# Cavity and HOM Coupler Design for CEPC*

ZHENG Hong-Juan (郑洪娟)[1] GAO Jie (高杰)[1] LIU Zhen-Chao(刘振超)[1]

1 (Key Laboratory of Particle Acceleration Physics and Technology, Institute of High Energy Physics, CAS, Beijing 100049, China)

**Abstract:** In this paper we will show a cavity and higher order mode (HOM) coupler designing scheme for the Circular Electron-Positron Collider (CEPC) main ring. The cavity radio frequency (RF) design parameters are showed in this paper. The HOM power is calculated based on the beam parameters in the Preliminary Conceptual Design Report (Pre-CDR). The damping results of the higher order modes (HOMs) and same order modes (SOMs) show that they are reached the damping requirements for beam stability.

## 1 Introduction

With the discovery of Higgs boson on LHC in 2012, the world's high energy physics (HEP) community is interested in future large circular colliders to study the Higgs boson. Because the Higgs mass is low (126 GeV), a circular $e+e-$ collider can serve as a Higgs factory. The Institute of High Energy Physics (IHEP) in Beijing, in collaboration with a number of other institutes, launched a study of 50-100 km ring collider [1]. It will serve as an $e+e-$ collider for the Higgs factory with the name of Circular Electron-Positron Collider (CEPC). A Preliminary Conceptual Design Report (Pre-CDR) was published on March, 2015 [2].

The circumference of the main ring is 54.7 km with 2 beams in one ring. The synchrotron radiation power for one beam is 51.7 MW. Table 1 shows the main parameters of the CEPC main ring [2]. The radio frequency (RF) system accelerates the electron and positron beams, compensates the synchrotron radiation loss and provide sufficient RF voltage for energy acceptance and the required bunch length in the collider. Superconducting Radio Frequency (SRF) cavities are used because they have much higher continuous wave (CW) gradient and energy efficiency as well as large beam aperture compared to normal conducting cavities. CEPC will use 384 five-cell 650 MHz cavities for the main ring. The cavity design and HOM power analysis are shown in this paper.

## 2 Cavity design

The geometry of the superconducting cavity is shown in Fig. 1. The meaning and the effects of the cavity parameters are summarized in the following:

1) $R_{iris}$ is the radius of the iris. Fig. 2 (a) shows that large $R_{iris}$ is good for the cell-to-cell coupling. However, large $R_{iris}$ leads to large $E_p/E_{acc}$ and $H_p/E_{acc}$ value. Large $R_{iris}$ value also decreases the impedance of the fundamental mode.

2) $D$ is the radius of the equator. It is used for the frequency tuning.

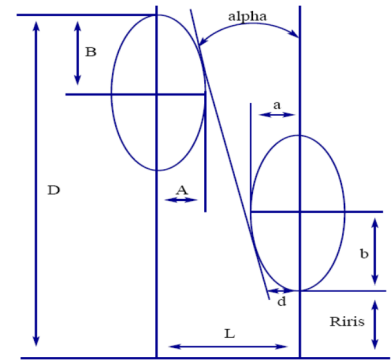

Fig. 1. The definition of cell shape parameters.

3) $L$ is the half length of the cavity cell. It is determined by the fundamental frequency and the cavity geometry beta.

4) The equator ellipse ration ($B/A$) is ruled by the mechanical considerations.

5) The iris ellipse ration ($b/a$) is determined by the local optimization of the peak electric field. Fig. 2 (b) shows that the iris ellipse ration is sensitive to $E_p/E_{acc}$ and cell-to-cell coupling. It has a little effect on $H_p/E_{acc}$ and $R/Q$.

6) *alpha* is the side wall inclination. Fig. 2 (c) shows that the angle has a little effect on $R/Q$. Small *alpha* decreases the $H_p/E_{acc}$ value while increases the cell-to-cell coupling. However, small *alpha* increases the $E_p/E_{acc}$ value. The higher angle is better for the cavity chemistry and cleaning procedures.

7) $d$ is the wall distance parameter. Fig. 2 (d) shows when $d$ increases, $E_p/E_{acc}$ will decrease while $H_p/E_{acc}$ will increase. The parameter $d$ is used to balance the $E_p/E_{acc}$ and $H_p/E_{acc}$ value. However, smaller $d$ also increases the cell-to-cell coupling.

* Supported by National Natural Science Foundation of China (11175192).
1) E-mail: zhenghj@ihep.ac.cn

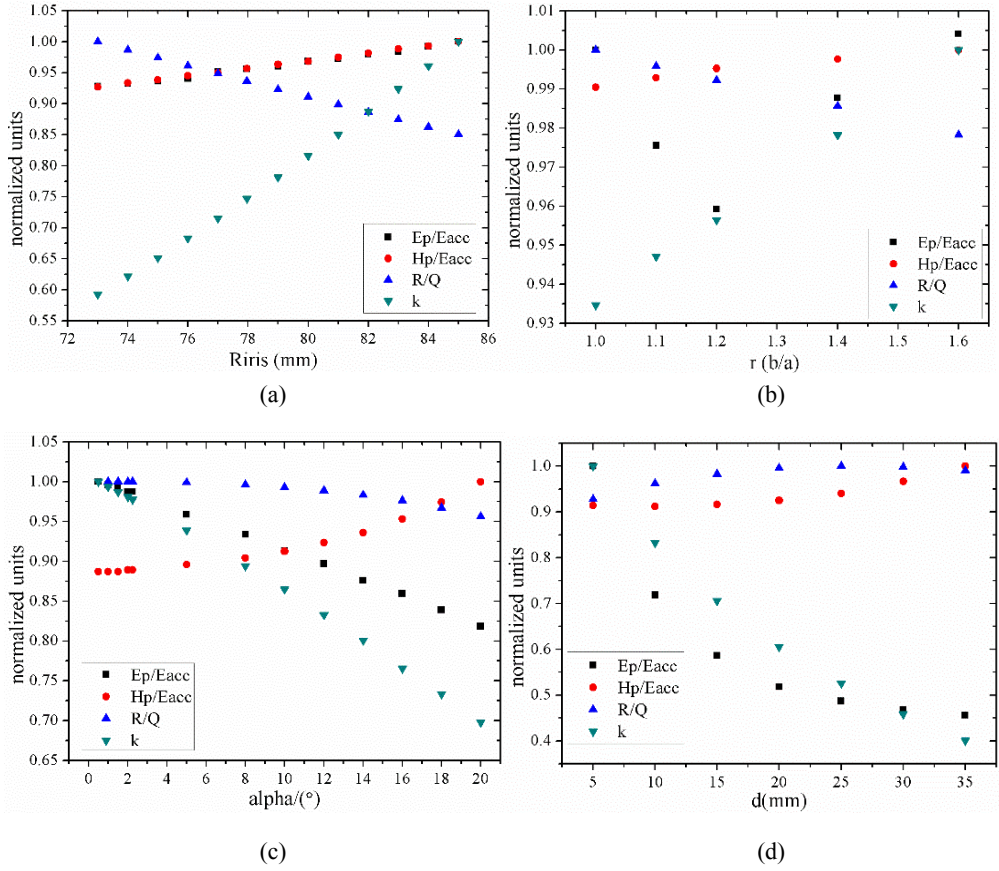

Fig. 2. Electromagnetic parameters as a function of $R_{iris}$, $b/a$, $alpha$, $d$.

Table 1. Main parameters for CEPC main ring.

| Parameter | Unit | Value | Parameter | Unit | Value |
|---|---|---|---|---|---|
| Beam energy | GeV | 120 | Circumference | m | 54752 |
| Number of IP | | 2 | SR loss/turn | GeV | 3.11 |
| Bunch number/beam | | 50 | Bunch population | | 3.79E+11 |
| Synchrotron radiation (SR) power/beam | MW | 51.7 | Beam current | mA | 16.6 |
| Bending radius | m | 6094 | Momentum compaction factor | | 3.36E-5 |
| Revolution period | s | 1.83E-4 | Revolution frequency | Hz | 5475.46 |
| Emittance (x/y) | nm | 6.12/0.018 | $\beta_{IP}$ (x/y) | mm | 800/1.2 |
| Transverse size (x/y) | μm | 69.97/0.15 | $\xi_{x,y}$/IP | | 0.118/0.083 |
| Bunch length SR | mm | 2.14 | Bunch length total | mm | 2.65 |
| Lifetime due to Beamstrahlung | min | 47 | Lifetime due to radiation Bhabha scattering | min | 51 |
| RF voltage | GV | 6.87 | RF frequency | MHz | 650 |
| Harmonic number | | 118800 | Synchrotron oscillation tune | | 0.18 |
| Energy acceptance RF | % | 5.99 | Damping partition number | | 2 |
| Energy spread SR | % | 0.132 | Energy spread BS | % | 0.096 |
| Energy spread total | % | 0.163 | During the collision | | 0.23 |
| Transverse damping time | turns | 78 | Longitudinal damping time | turns | 39 |
| Hourglass factor | Fh | 0.68 | Luminosity/IP | cm$^{-2}$s$^{-1}$ | 2.04E+34 |

We use the code Buildcavity [3] and Superfish [4] to do the parameter san analysis. The optimized parameters of the cavity based on the analysis above are shown in Table 2. The cell shape parameters are shown in Table 3.

The asymmetric end cell design is better to extract the HOMs. The cut-off frequency of the waveguide modes for the beam pipe are 1.355 GHz (TM01) and 1.04 GHz (TE11).

Table 2. The 650 MHz superconducting cavity parameters.

| Parameter | Value |
| --- | --- |
| Operating frequency/ MHz | 650 |
| No. of cells | 5 |
| Cavity effective length/ mm | 1147 |
| Cavity iris diameter/ mm | 156 |
| Beam tube diameter/ mm | 170 |
| Cell-to-cell coupling | 3% |
| $R/Q$/ Ω | 514 |
| Geometry factor/ Ω | 268 |
| $E_p/E_{acc}$ | 2.4 |
| $(H_p/E_{acc})$ / (mT/(MV/m)) | 4.23 |
| Acceptance gradient/ MV/m | 20 |
| Acceptance $Q_0$ | 4E10 |

Table 3. The 650 MHz superconducting cavity cell shape parameters.

| Parameters | Mid-cell | Left end cup | Right end cup |
| --- | --- | --- | --- |
| $L$/ mm | 115 | 114 | 113 |
| $R_{iris}$/ mm | 78 | 84.5 | 84.5 |
| $A$/ mm | 94.4 | 92.1 | 91 |
| $B$/ mm | 94.4 | 92.1 | 91 |
| $a$/ mm | 20 | 13.8 | 13.7 |
| $b$/ mm | 22.1 | 21.1 | 20.3 |
| $D$/ mm | 206.6 | 206.6 | 206.6 |
| alpha/ (°) | 2.24 | 16.7 | 16.4 |

## 3 HOM power

The bunch length of the main ring is 2.65 mm with a bunch population of 3.78E11. The total longitudinal loss factor of the cavity can be calculated by the program ABCI [5]. The loss factor of the HOMs and the total HOM power can be get by

$$k_0 = \frac{\omega_0}{4} \cdot \frac{R}{Q} \cdot e^{-\omega_0^2(\sigma_z/c)^2} \quad (1)$$

$k_0$ is the loss factor of the fundamental mode. $\omega_0$ is the frequency of the fundamental mode. $R/Q$ is the ratio of the transverse shunt impedance to its quality factor for the fundamental mode. $\sigma_z$ is the bunch length. $c$ is the speed of light.

$$k_{HOM} = k_t - k_0 \quad (2)$$

$k_{HOM}$ is the loss factor of the HOMs. $k_t$ is the total loss factor from ABCI.

$$P_{HOM} = 2k_{HOM} \cdot Q_b \cdot I_0 \quad (3)$$

$P_{HOM}$ is the HOM power. $Q_b$ is the bunch charge. $I_0$ is the beam current. 2 means two beam in one ring. According to the simulation results, the loss factor of the HOMs is 1.8 V/pC. The HOM power for each cavity is 3.6 kW. Below the cut-off frequency ($f$=1.355 GHz) of the beam pipe, all power should be extracted by the HOM coupler. Fig. 3 shows the distribution of power from the cavity HOM properties.

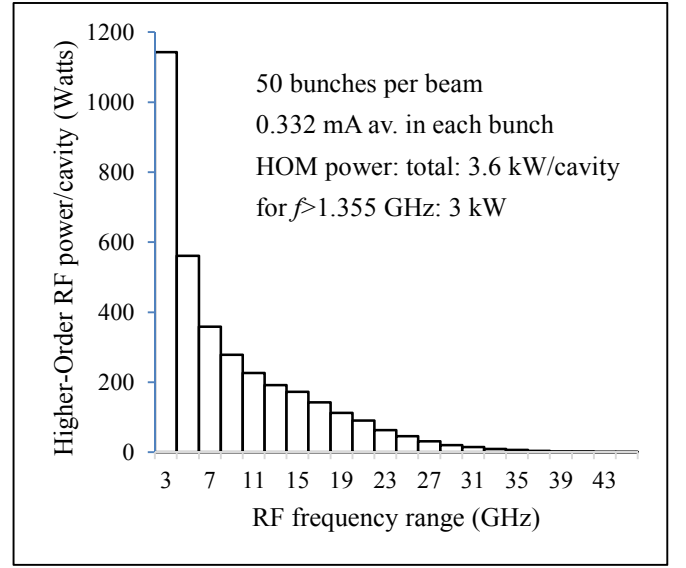

Fig. 3. Frequency distribution of HOM power.

## 4 SOMs and HOMs damping
### 4.1 SOMs damping

There are other four passband modes of the operating mode which are called the Same Order Modes (SOMs). Since the SOMs are close in frequency to the operating mode, they can't be damped in the same way as the HOMs. These modes may drive instabilities or extract significant RF power from the beam. We consider to use the input coupler as the SOM coupler. The external quality factor ($Q_e$) of the input coupler for the fundamental mode is 2.2E6. Both 2-D model and 3-D model are used to calculate the $Q_e$ of SOMs. The results are shown in Table 4.

In a storage ring, the beam instabilities in both the longitudinal and the transverse directions caused by a RF system are mainly from the cavities. To keep the beam stable, the radiation damping time should be less than the rise time of the multi-bunch instability. The threshold for the longitudinal impedance is given by [6]

$$R_L^{thresh} = \frac{2(E_0/e)\nu_s}{N_c f_L I_0 \alpha_P \tau_z} \quad (4)$$

$E_0$ is the beam energy. $e$ is the electronic charge. $\nu_s$ is the synchrotron oscillation tune. $N_c$ is the total number of cavity. $f_L$ is the mode frequency. $\alpha_p$ is the momentum compaction factor. $\tau_z$ is the longitudinal damping time. Bringing the beam parameters from Table 1 into this equation, the longitudinal impedance threshold for different modes can be get.

The threshold of the external quality factor of the mode is given by

$$Q_{\text{limit}} = \frac{R_L^{\text{thresh}}}{R/Q} \quad (5)$$

According to the equation (4) and (5), the $Q_{\text{limit}}$ results are also shown in Table 4.

If we consider the worst case, all the modes are on resonance ($\Delta\omega T_b=0$, $T_b << T_d$) [7]. The power dissipated by the beam is

$$P_{b,n} = \frac{R_a I_0^2}{1+\beta} \quad (6)$$

$$\beta = \frac{Q_0}{Q_e} \quad (7)$$

$R_a$ is the shunt impedance of the mode. $\beta$ is the coupling factor. $Q_0$ is the unloaded quality factor. $T_b$ is the bunch spacing time. $T_d$ is the decay time of the mode.

If we consider the real cavity passband modes frequencies and the bunch time spacing of the collider ($\Delta\omega T_b \neq 0$), the power dissipated by the beam is

$$\omega_n = \omega + \Delta\omega \quad (8)$$

$$\omega = \frac{2\pi h}{T_b} \quad (9)$$

$$F_r = \frac{1 - \exp(-2\frac{T_b}{T_d})}{2[1 - 2\exp(-\frac{T_b}{T_d})\cos(\Delta\omega T_b) + \exp(-2\frac{T_b}{T_d})]} \quad (10)$$

$$T_d = \frac{2Q_L}{\omega_n} = \frac{2Q_0}{\omega_n(1+\beta)} \quad (11)$$

$$P_{b,n} = \frac{R_a I_0^2}{1+\beta} \frac{T_b}{T_d} F_r \quad (12)$$

$\omega_n$ is the frequency of the SOM. The integer $h$ is the harmonic number of the beam and $\omega$ is the frequency that governs the bunch spacing. $Q_L$ is the loaded quality factor.

The power dissipated by the beam can be get by the analysis above. The results are shown in Table 4. $P_{\text{SOM-res}}$ stands for the worst case and $P_{\text{SOM}}$ stands for the general case. The analysis results show that the total SOM power is quite small when we consider the real cavity passband modes frequencies and the bunch time spacing of the collider. Even assuming resonant excitation, the total SOM power is about 1 kW and with the input coupler damping, the power dissipated on the cavity wall is negligible (~ 0.1 W). In other words, the damping for the SOMs by the input coupler are very efficient.

### 4.2 HOMs damping

Higher order modes excited by the intense beam bunches must be damped to avoid additional cryogenic loss and multi-bunch instabilities. This is accomplished by extracting the stored energy via HOM couplers mounted on both sides of the cavity beam pipe and the HOM absorbers outside the cryomodule.

Table 4. SOMs damping of the 650 MHz 5-cell cavity by the input coupler.

| Mode | $f$ (MHz) | $R/Q$ ($\Omega$) | $Q_{\text{limit}}$ | $Q_e$ | $P_{\text{SOM}}$ (W) | $P_{\text{SOM-res}}$ (W) |
|---|---|---|---|---|---|---|
| π/5 | 632.322 | 0.02 | 4.5E+9 | 1.2E+07 | 1.3E-5 | 268.9 |
| 2π/5 | 637.099 | 0.00017 | 5.4E+11 | 3.3E+06 | 8.7E-7 | 0.6 |
| 3π/5 | 643.139 | 0.341 | 2.6E+8 | 1.7E+06 | 9.31E-3 | 638.9 |
| 4π/5 | 648.146 | 0.078 | 1.1E+9 | 1.2E+06 | 2.92E-4 | 105.8 |

There are three major varieties of HOM couplers mainly used: waveguide, coaxial and beam tube. The coaxial HOM coupler is prevalent on several SRF cavity designs, such as XFEL, ILC, and SNS. There have challenges revealed with loop couplers in operating machines, such as with antenna overheating, multipactor, and mis-tuning due to fabrication variations. Beam tube couplers use large diameter beam pipes so as to reduce the R/Q values of the HOMs, as well as to propagate the HOMs freely out of the cavity to high power located at room temperature. This type of couplers will decrease the accelerating efficiency. They are mainly used for high current applications, such as KEK-B and CESR-III. The waveguide coupler has the benefit of transporting the HOM power to external room temperature loads. The cut-off is a natural rejection filter for the fundamental mode. It has the high power handling capability. The structure of waveguide coupler is simple. The waveguide coupler has been used in several facilities, such as PEPII, CEBAF and JLab high current cryomodule design [8].

The waveguide couplers are chose for HOMs damping. The high-pass filtering characteristics of waveguide is used to make the operating frequency of cavity under the cut-off frequency of waveguide. The higher order modes can propagate from the waveguide. We choose the cut-off frequency a little lower than the first higher order mode as the cut-off frequency of the waveguide. There are five waveguide HOM couplers and one coaxial main coupler spaced quite symmetrically to minimize transverse kick to the beam. The angular arrangement takes care of all mode polarizations. The five cell cavity with waveguide couplers is shown in Fig. 4.

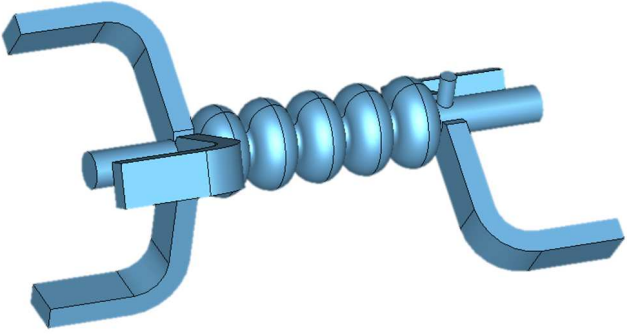

Fig. 4. 5 cell cavity with waveguide HOM couplers.

The higher order modes of the accelerating cavities will lead to the coupled bunch instability. The threshold for the longitudinal impedance is given by equation (4). The transverse coupled bunch instability is mainly caused by the dipole modes in the cavity. The threshold for the transverse impedance can be expressed as

$$R_T^{thresh} = \frac{2(E_0/e)}{N_c f_{rev} I_0 \beta_{x,y} \tau_{x,y}} \quad (13)$$

where $f_{rev}$ is the revolution frequency, $\beta_{x,y}$ is the beta function in the RF cavity region, and $\tau_{x,y}$ is the transverse damping time.

The HOMs are calculated up to 2 GHz. The simulation results compared with the impedance threshold are shown in Fig. 5 and Fig. 6. Fig. 5 shows the damping results for the monopole modes. As can be seen from the results, the damping for all the monopole modes are under the longitudinal impedance threshold. The $Q_e$ for fundamental mode is 4E11. The $Q_e$ for most of monopole modes is below $10^3$ while only a few of them is above $10^5$. The HOM power loss on the cavity is only a few milliwatts if the $Q_0$ of the monopole mode is $10^{10}$. Fig. 6 shows the damping results for the dipole modes. Most of the dipole modes damping results are under the transverse impedance threshold. However, we didn't take into account the spread in the resonance frequencies of different cavities. If the frequency spread is 0.5 MHz, the impedance threshold can increase 1~2 orders of magnitude [9]. So, this design can meet the requirements for beam stability.

## 5 HOM check

Two separate eigenmode simulations for just a single cell, with periodic boundary conditions (PBC) [10], were computed. One can control the phase shift from one cell to the other when using the PBC. In the end, 34 modes up to 1.8 GHz were calculated. One can get the resonant frequency of the fundamental 0-mode $f_0$ (0° phase shift at PBC) and the π–mode $f_\pi$ (180° phase shift at PBC). The same rules apply to the higher order modes. To calculate the cell-to-cell coupling factor one needs only $f_0$ and $f_\pi$

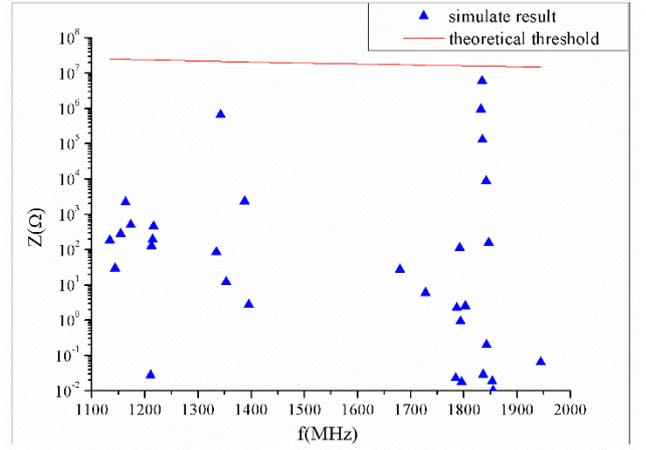

Fig. 5. Monopole modes damping results compared with impedance threshold.

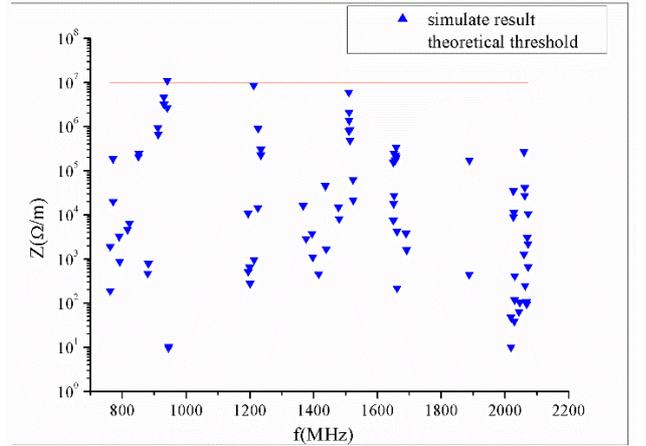

Fig. 6. Dipole modes damping results compared with impedance threshold.

$$k_{cc} = 2 \cdot \frac{f_\pi - f_0}{f_\pi + f_0} \quad (14)$$

The factor $k_{cc}$ specifies the passband width and smaller $k_{cc}$ gives narrower passband. For small ($|k_{cc}| \leq 0.01$) values, there is a danger that if the given mode is excited by the beam, e.g., somewhere in the middle of the cavity, it will propagate out and decay very slowly. Thus the factor $k_{cc}$ give us a preliminary knowledge of which modes can be dangerous or trapped. The results of the eigenmode analysis are listed in Table 5, including the $f_0$ and $f_\pi$ of all the modes, mode type, cell-to-cell coupling, and mode impedance. There are two modes with small $k_{cc}$ values, one of which is TE M3 mode. And the other one is TM D5 mode. After the field pattern analysis, the TE M3 mode is TE011 mode and the TM D5 mode is TM120 mode. There is no longitudinal electric field on axis for monopole mode TE012. The impedance for the dipole mode TM120 is also under the impedance threshold ($10^7$ Ω/m). Although the cell-to-cell coupling factors of the two modes are small, they are no dangerous.

## 6 Conclusion

In this paper, we give a reasonable cavity and HOM coupler design scheme for the CEPC main ring. The cavity parameters are given in the paper. The frequency distribution of the HOM power shows that the total HOM power for each cavity is 3.6 kW. After the waveguide HOM coupler introduced to the cavity, all the HOMs are under the impedance threshold. The SOMs are also damped successful by the input coupler. All of the analysis results show that the damping scheme are successful.

Table 5. Eigenmode analysis results.

| Phase advance 0° | | | Phase advance 180° | | | | |
|---|---|---|---|---|---|---|---|
| mode | $f$ (MHz) | $Z=(R/Q)*Q_e$ | mode | $f$ (MHz) | $Z=(R/Q)*Q_e$ | type | $k_{cc}$ |
| 1 | 0.630 | 1.21E+11 | 1 | 0.650 | 9.15E+13 | TM M1 | 0.0303 |
| 2,3 | 0.759 | 1.91E+3 | 4,5 | 0.887 | 2.46E+5 | TE D1 | 0.1551 |
| 4,5 | 0.946 | 1.03E+1 | 2,3 | 0.877 | 8.00E+2 | TM D2 | -0.075 |
| 8 | 1.182 | 1.79E+2 | 8 | 1.130 | 5.02E+2 | TM M2 | -0.0466 |
| 9 | 1.213 | 2.7E-2 | 11 | 1.218 | 1.43E-3 | TE M3 | 0.0045 |
| 10,11 | 1.238 | 1.63E+4 | 16,17 | 1.361 | 1.67E+3 | TM D3 | 0.0948 |
| 14 | 1.331 | 8.48E+1 | 18 | 1.400 | 2.76 | TM M4 | 0.0506 |
| 17,18 | 1.479 | 5.93E+6 | 19,20 | 1.509 | 2.15E+4 | TE D4 | 0.0202 |
| 27,28 | 1.652 | 1.56E+5 | 29,30 | 1.661 | 4.26E+3 | TM D5 | 0.0057 |
| 35 | 1.802 | -------- | 31 | 1.665 | 2.69E+1 | TM M4 | -0.0791 |